\documentclass[preprint,showpacs,preprintnumbers,amsmath,amssymb,aps]{revtex4}

\usepackage{graphicx}% Include figure files
\usepackage{dcolumn}% Align table columns on decimal point
\usepackage{bm}% bold math
\usepackage{epsfig}

\begin{document}
%===============

%\preprint{APS/123-XXX}

\title{Heavy atom tunneling in chemical reactions: study of $\mbox{H}+\mbox{LiF}$ collisions}

\author{P. F. Weck}
\email{weckp@unlv.nevada.edu}
\author{N. Balakrishnan}
\email{naduvala@unlv.nevada.edu}

\affiliation{Department of Chemistry, University of Nevada Las Vegas, 
 4505 Maryland Parkway, Las Vegas, NV 89154, USA}

\date{\today}

%===============
\begin{abstract}

The $\mbox{H}+\mbox{LiF}(X ^1\Sigma^+,v=0-2,j=0)\rightarrow\mbox{HF}(X ^1\Sigma^+,v',j')+\mbox{Li}(^2S)$ 
bimolecular process is investigated by means of quantum scattering calculations on the chemically accurate 
$X~^2A'$ LiHF potential energy surface of \citeauthor{agua03} [J. Chem. Phys. \textbf{119}, 10088 (2003)]. 
Calculations have been performed for zero total angular momentum for translational energies from $10^{-7}$ 
to $10^{-1}~\mbox{eV}$.
Initial-state selected reaction probabilities and cross sections are characterized 
by resonances originating from the decay of metastable states of the 
$\mbox{H}\cdots\mbox{F}-\mbox{Li}$ and $\mbox{Li}\cdots\mbox{F}-\mbox{H}$ van der Waals complexes. 
Extensive assignment of the resonances has been carried out by performing quasibound states calculations 
in the entrance and exit channel wells. Chemical reactivity is found to be significantly enhanced by 
vibrational excitation at low temperatures, although reactivity appears much less favorable 
than non-reactive processes due to the inefficient tunneling of the relatively heavy fluorine atom 
strongly bound in van der Waals complexes.  
 
\end{abstract}
%=============

\pacs{33.70.-w}% PACS, the Physics and Astronomy
                             % Classification Scheme.
\maketitle

%=======================================
\section{Introduction \label{sec:intro}} 
%=======================================

Quantum mechanical tunneling through an energy barrier has long been recognized as an 
important process in a variety of chemical reactions \cite{bell80}.
Recently, evidences for tunneling in reactions have been reported, for instance, in 
enzymology \cite{gao02}, organic \cite{zuev03}, atmospheric \cite{sulta04},  
and interstellar chemistry \cite{hira01}.  
This mechanism is also being investigated at cold and ultracold temperatures 
in the context of chemical reactivity involving Bose-Einstein condensates (BEC) 
of diatomic molecules \cite{joc03,gre03,zwi03} where tunneling is thought to be the  
dominant reactive process \cite{bala01,bodo02,bala03,weck04}.

Analysis of subthreshold resonances 
in the energy dependence of reaction cross sections can provide important  
insight into the reaction mechanism as illustrated recently for FH$_2$ \cite{taka98,taka99}, 
HOCl \cite{zou01,xie02}, HCl$_2$\cite{bowman05} and LiHF \cite{pania99,wei03,weck04b}.
For these systems, resonances were attributed to the decay of quasibound van der Waals 
complexes in the entrance and/or exit channels of the reaction. 
The LiHF system is especially interesting since at 
low temperatures where tunneling is preponderant, the reaction pathway 
between the $\mbox{H}+\mbox{LiF}$ and $\mbox{Li}+\mbox{HF}$ arrangements involves 
the transfer of the relatively heavy F atom. 
The $\mbox{LiH}+\mbox{F}$ product channel is highly endoergic and is not accessible 
at low energies.
The LiHF system has been extensively studied experimentally and theoretically.
Indeed, after the original crossed beam measurement of \citet{tayl55}, the 
$\mbox{Li}+\mbox{HF}$ reaction became a prototype for experimental 
investigation of the ``harpoon'' mechanism \cite{hers66} and many measurements of the 
integral and differential reactive cross sections have been 
reported \cite{beck80,loes93,baer94,aoiz99,aoiz00,casa00,hobe01,aoiz01,hobe04}.
Moreover, LiHF is particularly amenable to {\em ab initio} calculations of chemical accuracy, 
therefore explaining the variety of potential energy surfaces (PESs) available for the $X^2A'$ symmetry 
electronic ground state \cite{zeir78,chen80,cart80,laga84,garc84,
pani89,palm89,suar94,agua95,agua97a,agua97b,laga98a,laga98b,burc00,jasp01,jasp02,
burc02,agua03}. A large number of quantum mechanical 
\cite{baer94,park95,gogt96,agua97a,agua97b,lara98,pani98,aoiz99,aoiz00,lara00,
aoiz01,wei03,laga04,weck04b} and classical trajectory \cite{aoiz00,aoiz00b} scattering 
calculations have been performed on these PESs for the $\mbox{Li}+\mbox{HF}$ collision. 
In contrast, the $\mbox{H}+\mbox{LiF}$ collision has received no theoretical 
and experimental attention, to the best of our knowledge. The lack of experimental results 
may be explained by the difficulty in working with metastable hydrogen atom sources in 
crossed beams experiments. 

In this work, we present quantum scattering calculations 
on the $X~^2A'$ LiHF electronic ground state for the
$\mbox{H}+\mbox{LiF}(X ^1\Sigma^+,v=0-2,j=0)\rightarrow\mbox{HF}(X ^1\Sigma^+,v',j')+\mbox{Li}(^2S)$ 
reaction. 
Calculations have been performed for zero total angular momentum using the recent 
high accuracy global PES of \citet{agua03}.
In the context of our study of the reverse reaction \cite{weck04b},   
particular effort is made here to assign resonances due to the decay 
of metastable states in van der Waals wells of the entrance and exit channels. 
The LiHF electronic ground state is typical of alkali metal 
halides with a saddle point resulting from the crossing of 
the $\mbox{Li}^++\mbox{HF}^-$ ionic state and a covalent configuration 
correlating to $\mbox{Li}(^2S)+\mbox{HF}(X ^1\Sigma^+)$. The strong 
dipole electric field of the products is at the origin of an exceptionally deep 
van der Waals minimum of about $0.24~\mbox{eV}$ for the 
$\mbox{Li}\cdots\mbox{F}-\mbox{H}$ complex, while the 
$\mbox{H}\cdots\mbox{F}-\mbox{Li}$ complex in the entrance channel is about 
$0.07~\mbox{eV}$ deep.
These van der Waals wells give rise to long-lived collision complexes and 
narrow scattering resonances in the energy dependence of reaction probabilities.
Since $\mbox{H}+\mbox{LiF}\rightarrow\mbox{HF}+\mbox{Li}$ involves the transfer of 
the relatively heavy F atom, it will be of particular interest to see 
whether the reaction will occur with significant rate coefficient at low 
energies and how resonances due to tunneling affect reaction rates. 

In Sec. II, we give a summary of the principal features of the PES 
of \citet{agua03}, followed by a concise review of quantum scattering theory 
and convergence tests assessing the validity of our numerical results.  
Initial-state-selected probabilities, cross sections, cumulative reaction probabilities 
as well as rate coefficients for both reactive and non-reactive open channels of the  
collision are presented in Sec. III. We also provide a detailed analysis of the resonances present 
in the energy depencence of cross sections by means of quasibound state calculations 
of the van der Waals complexes formed in the reagent and product channels.      
The effect of vibrational excitation on chemical reactivity at low temperatures and the 
contribution of resonances in reaction rates are discussed. Finally, a 
summary of our findings is given in Sec. IV.

%=====================
\section{Calculations}  \label{sec:1}
%=====================

%---------------------------------------------------------
\subsection{$X^2A'$ LiHF potential energy surface }  
%---------------------------------------------------------

The $X^2A'$ LiHF ground state PES of \citet{agua03} was used for the 
present scattering calculations. 
{\em Ab initio} calculations were performed for 6000 nuclear geometries at the 
multireference configuration interaction (MRCI) level of theory, using internally 
contracted wave functions including all single and double excitations and the Davidson 
correction (+Q).
The $\mbox{Li}^++\mbox{HF}^-$ and $\mbox{Li}^++\mbox{H}^-\mbox{F}$ ionic configurations, 
responsible for the curve crossing leading to the LiF products in the 
adiabatic $X^2A'$ state, were adequately described using 
a large atomic basis set. The crossing between the $\mbox{Li}^++\mbox{HF}^-$ ionic state 
and a covalent configuration correlating to $\mbox{Li}(^2S)+\mbox{HF}(X ^1\Sigma^+)$ was 
found to produce a pronounced saddle point in the Born-Oppenheimer electronic ground 
state. The resulting analytic PES constructed using the modified many-body expansion 
of \citet{agua92} exhibits deep van der Waals wells of 
$0.2407~\mbox{eV}~(5.551~\mbox{kcal/mol})$ 
and $0.0686~\mbox{eV}~(1.582~\mbox{kcal/mol})$ corresponding to the 
$\mbox{Li}\cdots\mbox{F}-\mbox{H}$ and $\mbox{H}\cdots\mbox{F}-\mbox{Li}$ complexes, 
respectively. The saddle point and the $\mbox{Li}(^2S)+\mbox{HF}$ products asymptote 
are at $+0.0642~\mbox{eV}$ and $-0.1861~\mbox{eV}$, respectively, relative to the 
$\mbox{H}(^2S)+\mbox{LiF}(X ^1\Sigma^+)$ asymptote with $E=0$ corresponding 
to the bottom of the LiF potential. The $\mbox{LiH}(X ^1\Sigma^+)+\mbox{F}(^2P)$ products 
channel lies at $+3.3839~\mbox{eV}$ relative to the energy origin, thus this reaction 
channel is closed for the energy range considered in this study.
The PES of \citet{agua03} significantly improves on the widely used PES of \citet*{park95} 
constructed from a limited set of {\em ab initio} data. It is also markedly different 
from the recent {\em ab initio} PES of \citet{jasp01}, which exhibits a $+0.076~\mbox{eV}$ 
higher saddle point separating the $\mbox{H}\cdots\mbox{FLi}$ and 
$\mbox{Li}\cdots\mbox{FH}$ complexes and an additional saddle 
point at $+0.014~\mbox{eV}$ in the reagent valley.
Using the PES of \citet{agua03}, we found that, when the zero-point energies of 
reactants and products are included, the 
$\mbox{H}+\mbox{LiF}(v=0,j=0)\rightarrow\mbox{Li}+\mbox{HF}(v'=0,j'=0)$ reaction has 
an endoergicity of $0.0112~\mbox{eV}~(0.2588~\mbox{kcal/mol})$, while for 
vibrationally excited $\mbox{LiF}(v=1,j=0)$ and $\mbox{LiF}(v=2,j=0)$ reagents the 
reaction producing $\mbox{HF}(v'=0,j'=0)$ becomes exoergic by 
$0.0998~\mbox{eV}~(2.3011~\mbox{kcal/mol})$ and $0.2089~\mbox{eV}~(4.8170~\mbox{kcal/mol})$, 
respectively.

%-------------------------------------------
\subsection{Quantum scattering calculations}  
%-------------------------------------------

Quantum reactive scattering calculations were carried out using the coupled-channel 
hyperspherical coordinate method as implemented in the ABC program \cite{sko00}. 
In this method, the Schr\"{o}dinger equation for the motion of three nuclei on the parametric 
representation of a single Born-Oppenheimer PES is solved in Delves hyperspherical 
coordinates with exact reactive scattering boundary conditions. 
For all arrangements of the collision products, parity-adapted $S-$matrix 
elements, $S^{J,P}_{v'j'k',vjk}$, are calculated for each $(J,P,p)$ triple. Here, 
$J$ is the total angular momentum quantum number, $P$ and $p$ are the triatomic and diatomic 
parity eigenvalues, respectively; $v$ and $j$ are the usual diatomic vibrational 
and rotational quantum numbers and $k$ is the helicity quantum number for 
the reactants, their primed counterparts referring to the products. After 
transformation of the parity-adapted $S-$matrix elements into their standard 
helicity representation, $S^{J}_{v'j'k',vjk}$, initial state selected cross sections 
are computed as a function of the kinetic energy, $E_{kin}$, according to
\begin{equation}\label{stscs}
\sigma_{vj}(E_{kin}) = \frac{\pi}{k^2_{vj}(2j+1)}\sum^{J_{max}}_{J=0} 
(2J+1) \sum_{v'j'k'k} \vert S^{J}_{v'j'k',vjk}(E_{kin})\vert ^2,
\end{equation}
where $k_{vj}$ is the incident channel wave vector and the helicity quantum numbers 
$k$ and $k'$ are restricted to the ranges $0\leqslant k \leqslant \mbox{min}(J,j)$ and 
$0\leqslant k' \leqslant \mbox{min}(J,j')$. For zero total molecular 
angular momentum and $s-$wave scattering in the incident channel, Eq. (\ref{stscs}) merely 
reduces to a summation over the quantum number $v'$ and $j'$. 
The cumulative reaction reaction probability (CRP) is also computed 
using the formula   
\begin{equation}\label{crp}
 N_{J=0}(E) = \sum_{v'j'k',vjk} \vert S^{J=0}_{v'j'k',vjk}(E)\vert ^2,
\end{equation}
where $E$ is the total energy and the summation is over all open channels of the reactant 
and product molecules.
For $J=0$, reaction rate coefficients are further evaluated as the product of the cross section 
and the relative velocity, as a function of the translational temperature, $T=E_{kin}/k_B$, 
where $k_B$ is the Boltzmann constant. 

%-----------------------------
%\subsection{Convergence tests}  
%-----------------------------

Owing to the predominance of quantum tunneling in reactions involving energy 
barriers at low temperatures, reaction probabilities are generally small for 
translationally cold and ultracold collisions. 
Therefore, extensive convergence tests have been carried out for 
the initial-state-selected and state-to-state reaction probabilities of
the exothermic $\mbox{H}+\mbox{LiF}(v=1,2,j=0)\rightarrow\mbox{Li}+\mbox{HF}(v',j')$ 
reactions. Convergence of our scattering calculations was checked with respect to 
the maximum rotational quantum number, $j_{max}$, the cut-off energy, $E_{max}$  
that control the basis set size, the maximum value of the hyperradius, 
$\rho_{max}$, and the size of the log derivative propagation sectors, $\Delta{\rho}$. 

Converged reaction probability for LiF formation in $\mbox{H}+\mbox{LiF}(v=1,2,j=0)$ 
collisions were obtained over the range $10^{-5}-10^{-3}~\mbox{eV}$ using the 
values $\rho_{max}=25.0~\mbox{a.u.}$ and $\Delta{\rho}=0.01~\mbox{a.u.}$
The accuracy of our numerical results over this energy range was better than $10^{-10}$ 
using a large basis set size constrained with the values of $j_{max}=25$ and 
$E_{max}=2.5~\mbox{eV}$. Using the same values of $\rho_{max}$ and $\Delta{\rho}$ as above,
the size of the basis set was further optimized by adjusting the value of the cut-off 
energy.
In general, the reaction probabilities are more sensitive to the 
basis set parameters at low energies. We performed the basis set optimization 
at an incident kinetic energy of $10^{-5}~\mbox{eV}$. We found that 
the state-to-state reactions   
$\mbox{Li}+\mbox{HF}(v=1,2,j=0)\rightarrow\mbox{H}+\mbox{LiF}(v'=0,j')$ 
with $E_{max}=2.5~\mbox{eV}$ and $E_{max}=2.0~\mbox{eV}$ resulted in 
nearly identical values of the reaction probabilities. 
The basis set corresponding to $E_{max}=2.0~\mbox{eV}$ 
was composed of 634 basis functions.  
Based on these convergence tests, values of $\rho_{max}=25.0~\mbox{a.u.}$, 
$\Delta{\rho}=0.01~\mbox{a.u.}$, $j_{max}=25$ and $E_{max}=2.0~\mbox{eV}$ 
were adopted in the calculations reported hereafter.

%===============================
\section{Results and discussion}  \label{sec:2}
%===============================

Fig. \ref{fig1}~ shows the initial state-selected reaction probability for 
HF formation in $\mbox{H}+\mbox{LiF}(v=0-2,j=0)$ collisions as a function of 
the translational energy. For LiF reagent initially in the rovibrational 
ground state, the reaction leading to HF formation has an endoergicity of 
$0.0112~\mbox{eV}~(0.2588~\mbox{kcal/mol})$, thus explaining the presence of 
the kinetic energy threshold in the lower panel of Fig. \ref{fig1}. 
Probability for nonreactive collisions also exhibits a threshold near
$3\times 10^{-4}~\mbox{eV}~(0.007~\mbox{kcal/mol})$ associated with rotational
excitation to $\mbox{LiF}(v=0,j=1)$.
At low energies, the reaction probabilities are small, because quantum tunneling 
of the fluorine atom is the main reaction pathway. Vibrational excitation of LiF  
significantly enhances the reaction probability in the ultracold and cold regimes, 
for instance, an energy increase of $0.109~\mbox{eV}$ from LiF$(v=1,j=0)$ to 
LiF$(v=2,j=0)$ translates in a 350\% increase of the reaction probability in the 
zero temperature limit. 
Our results also show a dense resonance structure in the energy dependence of 
the reaction probability at low energies associated with quasibound states of the 
$\mbox{H}\cdots\mbox{F}-\mbox{Li}$ and $\mbox{Li}\cdots\mbox{F}-\mbox{H}$ van 
der Waals complexes. Assignment of these resonances will 
be presented further in this study.            

Cross sections for $s-$wave elastic scattering in $\mbox{H}+\mbox{LiF}(v=0-2,j=0)$ 
collisions are displayed in Fig. \ref{fig2} as a function of the incident kinetic energy. 
For low translational energies, only $s-$wave scattering is expected to play a significant
role \cite{kaj03}. For collisions involving excited LiF reagents, elastic cross sections 
are found to have similar values for kinetic energies below $5\times10^{-3}~\mbox{eV}$. 
The nearly identical values of the limiting elastic cross sections for $v=1$ and 2 suggest 
that vibrational excitation does not affect the isotropic part of the interaction potential.
In the zero-temperature limit, the scattering length can be expressed as a complex number, 
$a_{vj}=\alpha_{vj}-i\beta_{vj}$, where the real and imaginary parts are related to   
the elastic component of the $S-$matrix, $S^{el}_{vj}$, according to 
\begin{equation}\label{rscatl}
\alpha_{vj} = -\lim_{k_{vj}\to 0}\frac{\mbox{Im}(S^{el}_{vj})}{2k_{vj}},
\end{equation}
\begin{equation}\label{rscat2}
\beta_{vj} = \lim_{k_{vj}\to 0}\frac{1-\mbox{Re}(S^{el}_{vj})}{2k_{vj}},
\end{equation}
where $k_{vj}$ is the wave vector in the incident channel. We found $\alpha_{10}=+9.100~\mbox{\AA}$ 
and $\alpha_{20}=+9.159~\mbox{\AA}$ for $v=1$ and $v=2$, respectively. 
However, for numerical reasons, the Wigner's threshold law for inelastic reactive and nonreactive 
collisions, i. e. 
\begin{equation}\label{sigin}
\sigma^{in}_{vj} = \frac{4\pi\beta_{vj}}{k_{vj}},
\end{equation}
is preferred over Eq. (\ref{rscat2}) to evaluate the imaginary part of the scattering length. 
In Eq. (\ref{sigin}), $\sigma^{in}_{vj}$ is the sum of the nonreactive and reactive cross sections.
Using Eq. (\ref{sigin}), we found $\beta_{10}=2.365\times 10^{-4}~\mbox{\AA}$ and   
$\beta_{20}=3.478\times 10^{-3}~\mbox{\AA}$ for $v=1$ and $v=2$, respectively.
The total elastic cross section in the zero energy limit is given by
\begin{equation}\label{sigel}
\sigma^{el}_{vj} = 4\pi \left| a_{vj}\right|^2,
\end{equation}
which yields $\sigma^{el}_{20}=1054.05\times 10^{-16}~\mbox{cm}^2$ and 
$\sigma^{el}_{10}=1040.58\times 10^{-16}~\mbox{cm}^2$.  

Initial-state-selected cross sections for HF formation and for 
nonreactive scattering in $\mbox{H}+\mbox{LiF}(v=0-2,j=0)$ collisions are displayed 
in Fig. \ref{fig3} as a function of the incident translational energy. 
For $v=1$ and $v=2$, reaction cross sections reach the Wigner regime \cite{wig48}
for kinetic energies below $10^{-5}~\mbox{eV}$ in accordance with Eq. (\ref{sigin}). 
Nonreactive scattering dominates 
over the complete kinetic energy range 
investigated, with a $\mbox{LiF}/\mbox{HF}$ product branching ratio  
of 4 for $v=1$ and 15 for $v=2$ in the Wigner regime. 
For translational energies beyond $10^{-4}~\mbox{eV}$, reaction cross sections  
are characterized by resonant spikes due to metastable states of the 
$\mbox{H}\cdots\mbox{F}-\mbox{Li}$ and $\mbox{Li}\cdots\mbox{F}-\mbox{H}$ van der Waals 
complexes. These resonances tend to be washed out for reactions with vibrational excited 
reactants.    

Assignments of the resonances in the energy dependence of the reactive 
cross sections in $\mbox{H}+\mbox{LiF}(v=0-2,j=0)$ collisions 
are presented in Table \ref{tab1}. Bound- and quasi-bound states calculations have 
been performed for the $\mbox{H}\cdots\mbox{F}-\mbox{Li}$ van der Waals potentials 
correlating with the $\mbox{LiF}(v=0-2)$ manifold. 
The adiabatic potentials are obtained by constructing the matrix elements of the 
interaction potential in a basis set of the rovibrational levels of the LiF molecule 
and diagonalizing the resulting diabatic potentials as a function of the 
atom-molecule separation, $R$. The resonance energies and the corresponding 
wave functions are computed using the Fourier grid Hamiltonian (FGH) method \cite{mars89,bali92}.   
For constructing the adiabatic potentials, we used a 20-term Legendre expansion 
of the interaction potential, 25 angular orientations to project out the expansion 
coefficients, 17 Gauss-Hermite quadrature points for the vibrational wave functions 
and a grid of 2000 points in the atom-molecule separation ranging from $R=0.1-100~a_0$.
Similar calculations for the $\mbox{Li}\cdots\mbox{F}-\mbox{H}$ exit channel complexes have 
been carried out previously \cite{weck04b} and results for the van der Waals potential correlating 
with the $\mbox{HF}(v=0)$ manifold were reported. Total energy values given in 
Table \ref{tab1} are relative to separated $\mbox{H}+\mbox{LiF}$ system with energy 
origin at the bottom of the LiF potential. 
Eigenenergies of the reactant and product diatoms calculated using the FGH and ABC codes 
were found to be consistent within an error margin of $10^{-6}~\mbox{eV}$ for both 
$\mbox{H}+\mbox{LiF}$ and $\mbox{Li}+\mbox{HF}$ asymptotes.
The resonances in the $\mbox{H}\cdots\mbox{LiF}$ entrance channel 
van der Waals well are characterized by quantum numbers $(v,j,t)$, where 
$v$ and $j$ are the LiF vibrational and rotational quantum numbers, respectively, and 
$t$ refers to the $\mbox{H}-\mbox{LiF}(v,j)$ van der Waals stretching 
vibration; their primed counterparts refer to resonances originating from the 
$\mbox{Li}\cdots\mbox{HF}$ exit channel well. As shown in Table \ref{tab1}, the agreement between 
the FGH energy eigenvalues and the peak positions from our scattering calculations is 
excellent. For $v=0$, resonances are mostly due to the decay of $(v'=0,j',t')$ metastable 
states of the $\mbox{Li}\cdots\mbox{HF}$ exit channel van der Waals complex, with $j'$ and $t'$ ranging from 
relatively high $t'$-values for $j'=1$ to the lowest-lying stretching vibrational states of $j'=5$. 
Resonances attributed to the $\mbox{H}\cdots\mbox{LiF}(v=0,j=10-15,t=1-2)$ states also appear 
for reactants in this vibrational state. For $v=1$ and $v=2$, this tendency is reverted with 
a dominant imprint of resonances corresponding to low$-t$ quasibound states of the 
adiabatic potentials correlating with $j=2-12$ of the entrance channel 
$\mbox{H}\cdots\mbox{LiF}$ complex. For these 
vibrationally excited states of the reagents, only few high-lying stretching 
vibrational states of the $\mbox{Li}\cdots\mbox{HF}$ exit channel van der Waals complex give rise 
to resonances: $j'=6$ for $v=1$ and $j'=9$ for $v=2$. Although the exit channel van der Waals well is much deeper than   
the entrance channel well, the $\mbox{Li}\cdots\mbox{HF}$ exit channel complex dominates only for 
$v=0$, while for higher vibrational levels of the reagents, the entrance channel complex plays the 
major role and the signature of the exit channel quasi-bound states tends to disappear.      
It is also interesting to note that the resonances appear mostly in the reactive cross sections 
despite the fact that for $v=1$ and 2 it is the entrance channel van der Waals potential that supports 
the resonances. Thus, it appears that the long lived resonances decay mostly by vibrational pre-reaction of 
the quasibound states by tunneling of the F atom.

The cumulative reaction probability for HF formation in $\mbox{H}+\mbox{LiF}(v=0-2,j=0)$ 
collisions is depicted in Fig. \ref{fig4} as a function of the total energy. CRP is 
represented over an energy range of $0.1~\mbox{eV}$ starting from the total energy values 
of the $\mbox{LiF}(v,j=0)$ reagents, thus explaining the presence of thresholds at 
$0.0675~\mbox{eV}$, $0.1673~\mbox{eV}$ and $0.2764~\mbox{eV}$, respectively, for $v=0-2$. 
The resonance structure still subsists in the CRP, but becomes less sharper with increase 
in energy and vibrational excitation.

The $J=0$ contribution to the reaction rate coefficients for HF formation in 
$\mbox{H}+\mbox{LiF}(v=1,2,j=0)$ collisions is shown in Fig. \ref{fig5}. 
For $v=1$, the reaction rate coefficient reaches the Wigner regime for temperatures below 
$0.1~\mbox{K}$, with a finite value of 
$3.86\times 10^{-15}~\mbox{cm}^3~\mbox{s}^{-1}$ in the zero-temperature limit.
Vibrational excitation of HF to the $v=2$ level enhances reactivity by about a factor of 4 
in the cold and ultracold regimes.
Although rate coefficients appear to be small at low temperatures, the values are 
not negligible compared to those for reactions involving tunneling 
of atoms lighter than fluorine. This suggests that heavy atom tunneling may play a major 
role at low energies, as observed recently by \citet{zuev03} for carbon tunneling 
in ring expansion reactions of 1-methylcyclobutylhalocarbenes at 8 to 25 K.        
In Fig. \ref{fig5}, for $v=1$ and $v=2$, the reactivity rapidly increases beyond $3~\mbox{K}$, 
however, accurate prediction of rate coefficients for higher temperatures requires calculations 
for $J>0$ which is beyond the scope of this work.

%===================
\section{Conclusion}  \label{sec:conclu}
%===================

We have investigated the scattering dynamics of the  
$\mbox{H}+\mbox{LiF}(X ^1\Sigma^+,v=0-2,j=0)\rightarrow\mbox{HF}(X ^1\Sigma^+,v',j')+\mbox{Li}(^2S)$
reaction for zero total angular momentum. Using the coupled-channel hyperspherical coordinate method, 
probabilities, cross sections, and rate coefficients have been 
computed from the zero-temperature limit up to a translational energy of $0.1~\mbox{eV}$ using 
the most recent PES for the LiHF electronic ground state. 
Extensive calculations have been carried out for quasibound states in the entrance and exit channel van 
der Waals wells to characterize the rich resonance structure observed in our scattering calculations. 
We found that the resonances arise from the decay of metastable states of both 
$\mbox{H}\cdots\mbox{F}-\mbox{Li}$ and $\mbox{Li}\cdots\mbox{F}-\mbox{H}$ van der Waals complexes. 
In particular, the quasibound states of the deep $\mbox{Li}\cdots\mbox{F}-\mbox{H}$ van der Waals 
well were found to strongly imprint the resonance structure for reactions involving 
$\mbox{LiF}(v=0,j=0)$. For vibrationally excited reagents, however, resonances are mainly 
originating from tunneling of metastable states of $\mbox{H}\cdots\mbox{F}-\mbox{Li}$~ into 
the $\mbox{HF}(v'=0)$ product manifold.  
Consistent with our recent findings for the reverse reaction \cite{weck04b}, chemical reactivity for 
$\mbox{H}+\mbox{LiF}\rightarrow\mbox{HF}+\mbox{Li}$ is enhanced by vibrational excitation of the 
reagents at low and ultralow temperatures. In contrast to $\mbox{Li}+\mbox{HF}(v=1,j=0)$, 
nonreactive processes dominate over HF formation in $\mbox{H}+\mbox{LiF}(v=0-2,j=0)$ collisions.       
However, quasibound states of the entrance channel van der Waals potential do decay 
through vibrational pre-reaction by tunneling of the fluorine atom.
The results suggest that heavy atom tunneling at low temperatures may be more important 
than generally recognized, as observed recently by \citet{zuev03} for carbon tunneling 
in ring expansion reactions.

%======================
\begin{acknowledgments}
%======================

This work was supported by NSF grant PHYS-0245019, the Research Corporation and by the 
United States-Israel Binational Science Foundation.

%====================
\end{acknowledgments}
%====================

%=============
% Bibliography
%=============

%====================

\clearpage

%======
%Tables 
%======

\begingroup
\squeezetable
\begin{table}
\caption{\label{tab1} Assignment of the resonances in the energy dependence of the 
cross sections for HF formation in $\mbox{H}+\mbox{LiF}(v=0-2,j=0)$ collisions 
(energies in eV).}
\begin{ruledtabular}
\begin{tabular}{cccccccccccc}
  \multicolumn{1}{c}{}
& \multicolumn{2}{c}{Position\footnotemark[2]}
& \multicolumn{1}{c}{} 
& \multicolumn{3}{c}{$\mbox{H}\cdots\mbox{LiF}(v,j,t)$} 
& \multicolumn{1}{c}{} 
& \multicolumn{4}{c}{$\mbox{Li}\cdots\mbox{HF}(v',j',t')$} 
\\
\cline{2-3} \cline{5-7} \cline{9-12}
  \multicolumn{1}{c}{$v$\footnotemark[1]}
& \multicolumn{1}{c}{$E_{kin}$}
& \multicolumn{1}{c}{$E_{tot}$}
& \multicolumn{1}{c}{} 
& \multicolumn{1}{c}{$E_{v,j,t}$\footnotemark[3] } 
& \multicolumn{1}{c}{$j$\footnotemark[4]} 
& \multicolumn{1}{c}{$t$\footnotemark[5]}
& \multicolumn{1}{c}{} 
& \multicolumn{1}{c}{$E_{v',j',t'}$\footnotemark[3] } 
& \multicolumn{1}{c}{$v'$\footnotemark[6]} 
& \multicolumn{1}{c}{$j'$\footnotemark[7]} 
& \multicolumn{1}{c}{$t'$\footnotemark[8]}
\\
\hline
%$\mbox{H}+\mbox{LiF}(v=0,j=0)$ && & & & & & & & & \\
%\hline
%
%0& &0.0680 &&        &    &     &&  X      & x & x  & x  \\
0&0.0127 &0.0690 &&         &    &     && 0.0689  & 0 & 2  & 6  \\
0&0.0132 &0.0695 && 0.0692  & 10 & 1   && 0.0694  & 0 & 1  & 11 \\
0&0.0146 &0.0709 &&         &    &     && 0.0709  & 0 & 3  & 3  \\
0&0.0157 &0.0720 &&         &    &     && 0.0718  & 0 & 1  & 12  \\
0&0.0165 &0.0728 &&         &    &     && 0.0725  & 0 & 4  & 1  \\
0&0.0175 &0.0738 &&         &    &     && 0.0737  & 0 & 2  & 7  \\
0&0.0222 &0.0785 && 0.0787  & 12 & 1   && 0.0786  & 0 & 3  & 4  \\
0&0.0244 &0.0807 && 0.0809  & 13 & 1   && 0.0804  & 0 & 2  & 9  \\
0&0.0264 &0.0827 &&         &    &     && 0.0826  & 0 & 2  & 10  \\
0&0.0272 &0.0835 &&         &    &     && 0.0836  & 0 & 4  & 1  \\
0&0.0280 &0.0843 &&         &    &     && 0.0843  & 0 & 3  & 5  \\
0&0.0298 &0.0861 && 0.0868  & 14 & 1   &&         &   &    &     \\
0&0.0357 &0.0920 && 0.0922  & 15 & 2   && 0.0920  & 0 & 5  & 1    \\
0&0.0397 &0.0960 &&         &    &     && 0.0957  & 0 & 3  & 8  \\
0&0.0437 &0.1000 &&         &    &     && 0.0992  & 0 & 4  & 4  \\
0&0.0477 &0.1040 &&         &    &     && 0.1043  & 0 & 5  & 2  \\
1&0.0003 &0.1676 && 0.1675  & 2  & 2   &&         &   &    &     \\
1&0.0007 &0.1680 && 0.1679  & 6  & 2   &&         &   &    &     \\
1&0.0018 &0.1691 && 0.1690  & 3  & 2   &&         &   &    &     \\
1&0.0028 &0.1701 &&         &    &     && 0.1701  & 0 & 6  & 8   \\
1&0.0033 &0.1706 && 0.1705  & 4  & 2   &&         &   &    &     \\
1&0.0049 &0.1722 &&         &    &     && 0.1726  & 0 & 6  & 9   \\
1&0.0061 &0.1734 && 0.1733  & 8  & 1   &&         &   &    &     \\
1&0.0093 &0.1766 && 0.1766  & 9  & 1   &&         &   &    &     \\
1&0.0131 &0.1804 && 0.1802  & 10 & 1   &&         &   &    &     \\
1&0.0177 &0.1850 && 0.1856  & 11 & 1   &&         &   &    &     \\
1&0.0235 &0.1908 && 0.1896  & 12 & 1   &&         &   &    &     \\
2&0.0010 &0.2774 && 0.2772  & 6  & 1   &&         &   &    &     \\
2&0.0029 &0.2793 && 0.2794  & 7  & 1   &&         &   &    &     \\
2&0.0046 &0.2810 &&         &    &     && 0.2807  & 0 & 9  & 6   \\
2&0.0056 &0.2820 && 0.2822  & 8  & 1   &&         &   &    &     \\
2&0.0095 &0.2859 && 0.2854  & 9  & 1   &&         &   &    &     \\
2&0.0118 &0.2882 && 0.2889  & 10 & 1   && 0.2880  & 0 & 9  & 8   \\
2&0.0168 &0.2932 && 0.2936  & 11 & 1   &&         &   &    &     \\
2&0.0225 &0.2989 && 0.2985  & 12 & 1   &&         &   &    &     \\
\end{tabular}
\end{ruledtabular}
\footnotetext[1]{LiF vibrational quantum number.}
\footnotetext[2]{Position of the observed resonances. Total energy, $E_{tot}$, is relative to 
separated $\mbox{H}+\mbox{LiF}$ system with energy zero corresponding to the bottom of the LiF potential.}
\footnotetext[3]{Binding energies $E_{v,j,t}$ and $E_{v',j',t'}$ are calculated with the Fourier 
grid Hamiltonian method in the $\mbox{H}\cdots\mbox{LiF}$ and $\mbox{Li}\cdots\mbox{HF}$ 
van der Waals wells, respectively.} 
\footnotetext[4]{LiF rotational quantum number.}
\footnotetext[5]{Quantum number for the $\mbox{H}-\mbox{LiF}(v,j)$ van der Waals stretching vibration.}
\footnotetext[6]{HF vibrational quantum number.}
\footnotetext[7]{HF rotational quantum number.}
\footnotetext[8]{Quantum number for the $\mbox{Li}-\mbox{HF}(v',j')$ van der Waals stretching 
vibration.}

\end{table}
\endgroup

\clearpage

%================
% Figure captions
%================

\begin{figure} 
 \caption{\label{fig1} Initial state-selected probability for 
 HF formation (solid curve) and for non-reactive scattering (dashed curve) in 
 $\mbox{H}+\mbox{LiF}(v=0-2,j=0)$ collisions as a function of the total energy.}
\end{figure}

\begin{figure} 
\caption{\label{fig2} Elastic cross sections for $s-$wave scattering in  
 $\mbox{H}+\mbox{LiF}(v=0-2,j=0)$ collisions as a function of the 
 incident kinetic energy. Solid curve: $v=0$; dotted curve: $v=1$; dashed curve: $v=2$.}
\end{figure}

\begin{figure} 
\caption{\label{fig3} Cross sections for HF formation and nonreactive 
 scattering in $\mbox{H}+\mbox{LiF}(v=0-2,j=0)$ collisions as a function of the 
 incident kinetic energy. solid curve: HF product channel; dashed curve: nonreactive 
 scattering.}
\end{figure}

\begin{figure} 
\caption{\label{fig4} Cumulative reaction probability for HF formation 
 in $\mbox{H}+\mbox{LiF}(v=0-2,j=0)$ collisions as a function of the 
 total energy. Solid curve: $v=0$; dotted curve: $v=1$; dashed curve: $v=2$.}
\end{figure}

\begin{figure} 
\caption{\label{fig5} Temperature dependence of reaction rate 
 coefficients for HF formation and nonreactive vibrational quenching 
 in $\mbox{Li}+\mbox{HF}(v=1,2,j=0)$ collisions. }
\end{figure}

%================
% Figure captions
%================

\clearpage

\begin{figure*}
\begin{center}
\includegraphics[height=14cm,width=16cm]{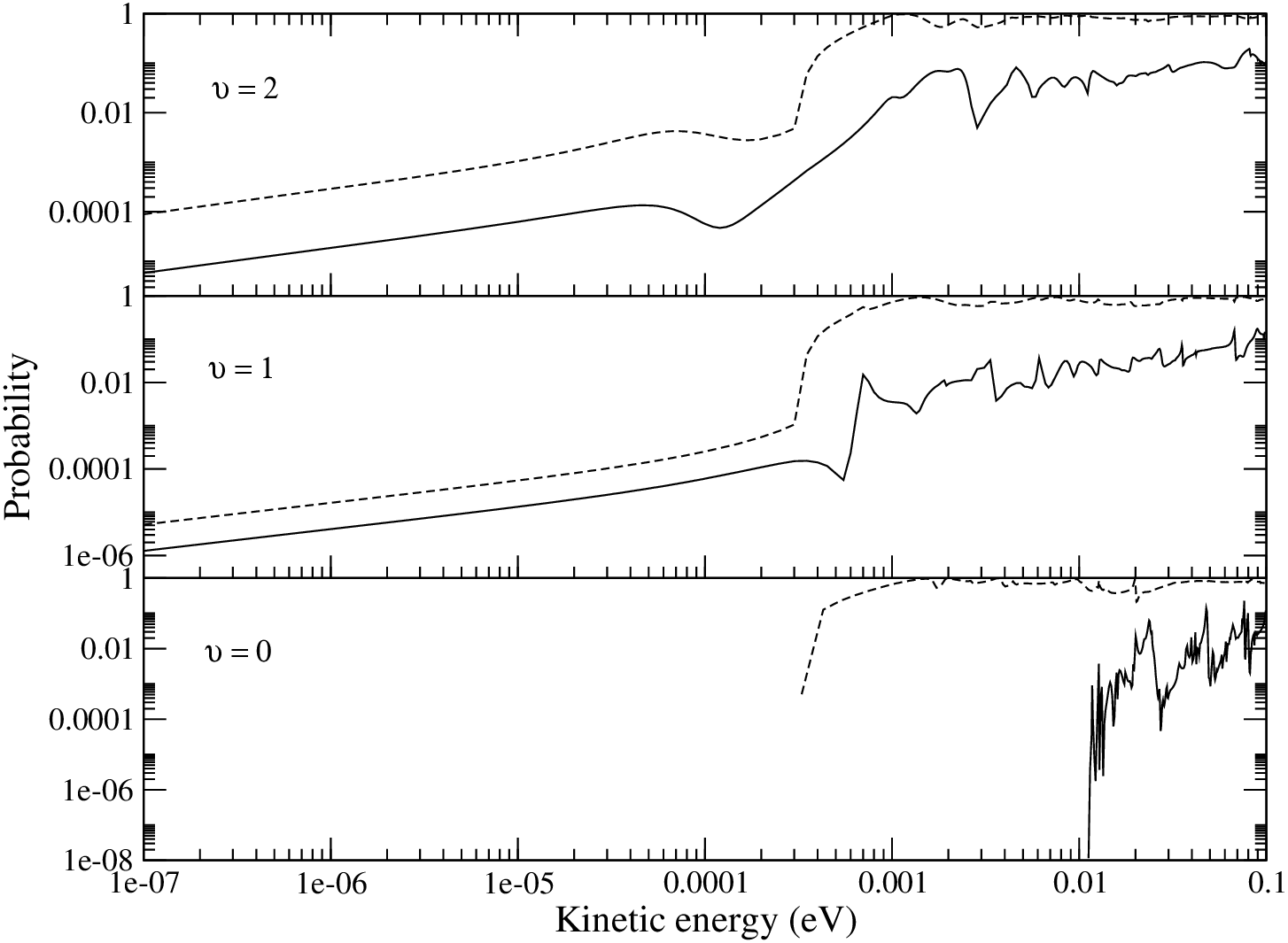}
\end{center} 
\end{figure*}

\clearpage
\begin{figure*}
\begin{center}
\includegraphics[height=14cm,width=16cm]{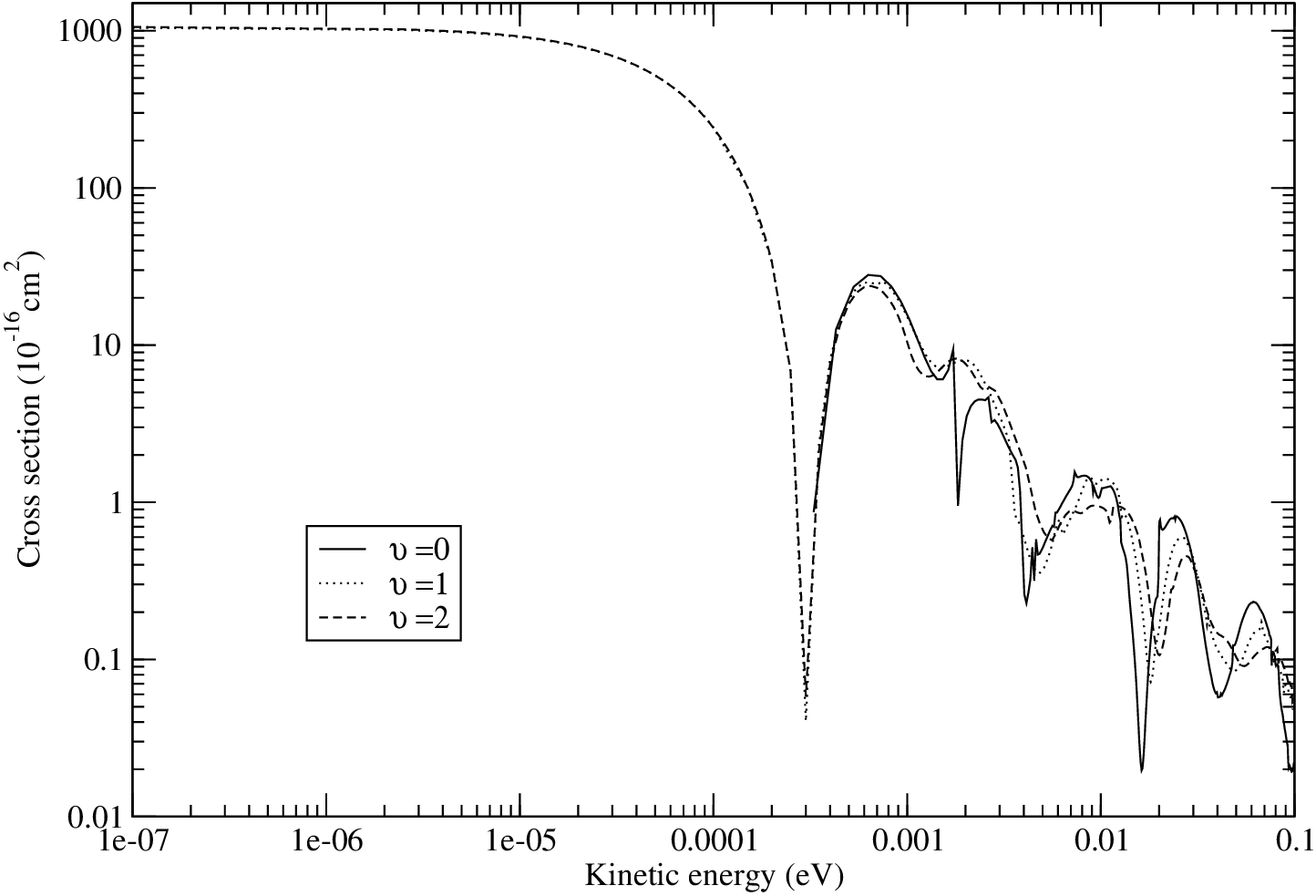}
\end{center} 
\end{figure*}

\clearpage
\begin{figure*}
\begin{center}
\includegraphics[height=14cm,width=16cm]{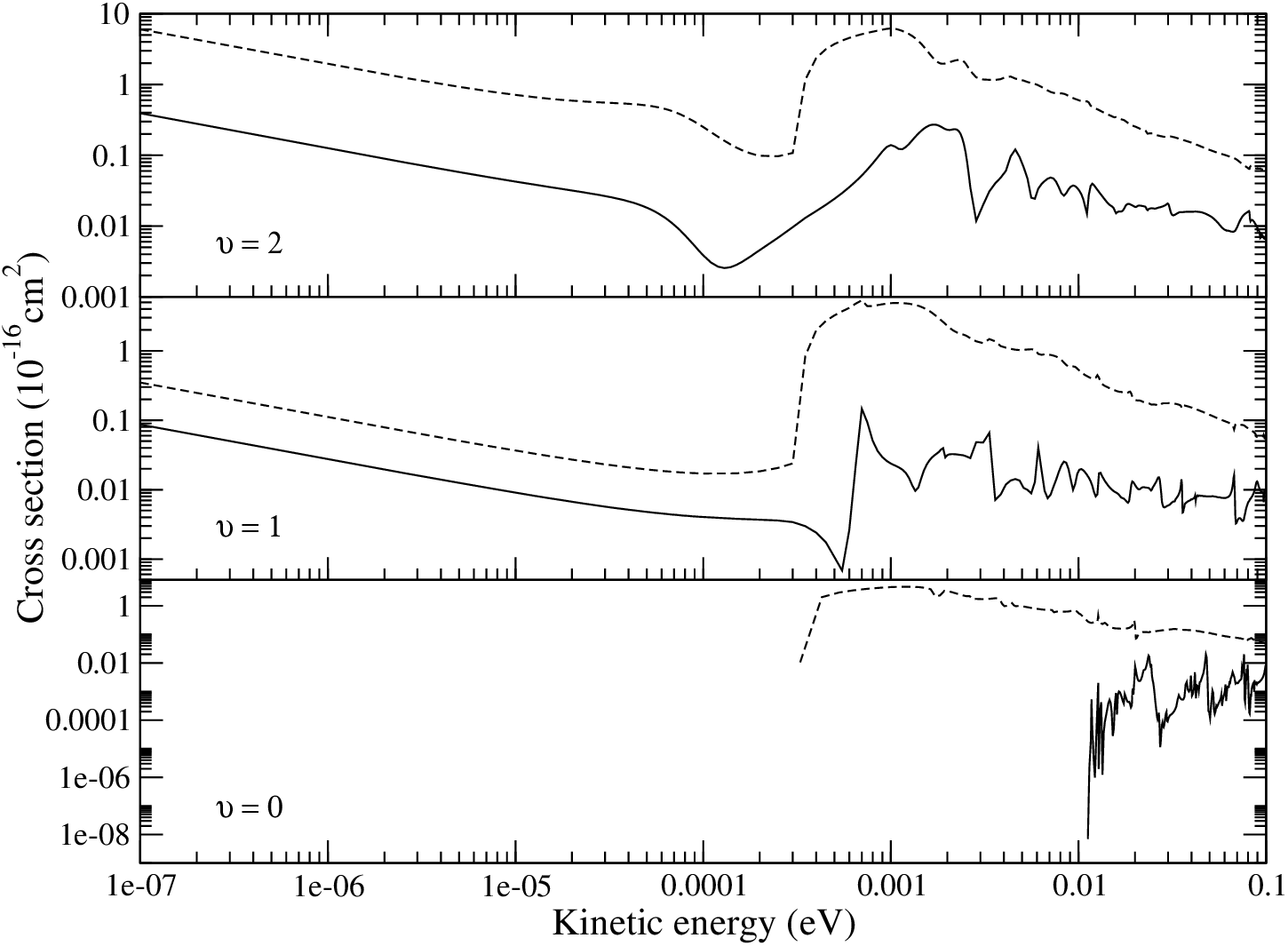}
\end{center} 
\end{figure*}

\clearpage
\begin{figure*}
\begin{center}
\includegraphics[height=14cm,width=16cm]{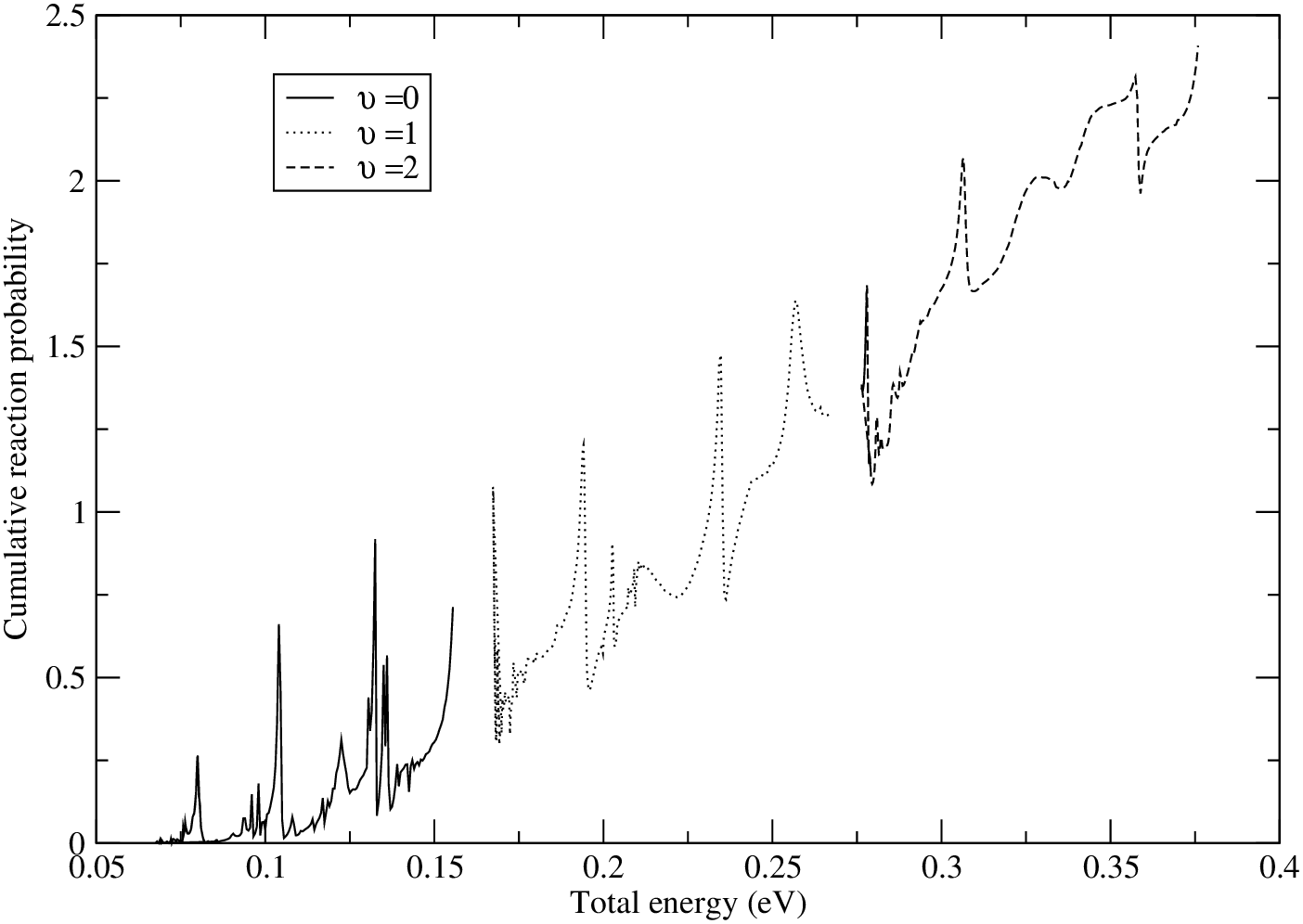}
\end{center} 
\end{figure*}

\clearpage
\begin{figure*}
\begin{center}
\includegraphics[height=14cm,width=16cm]{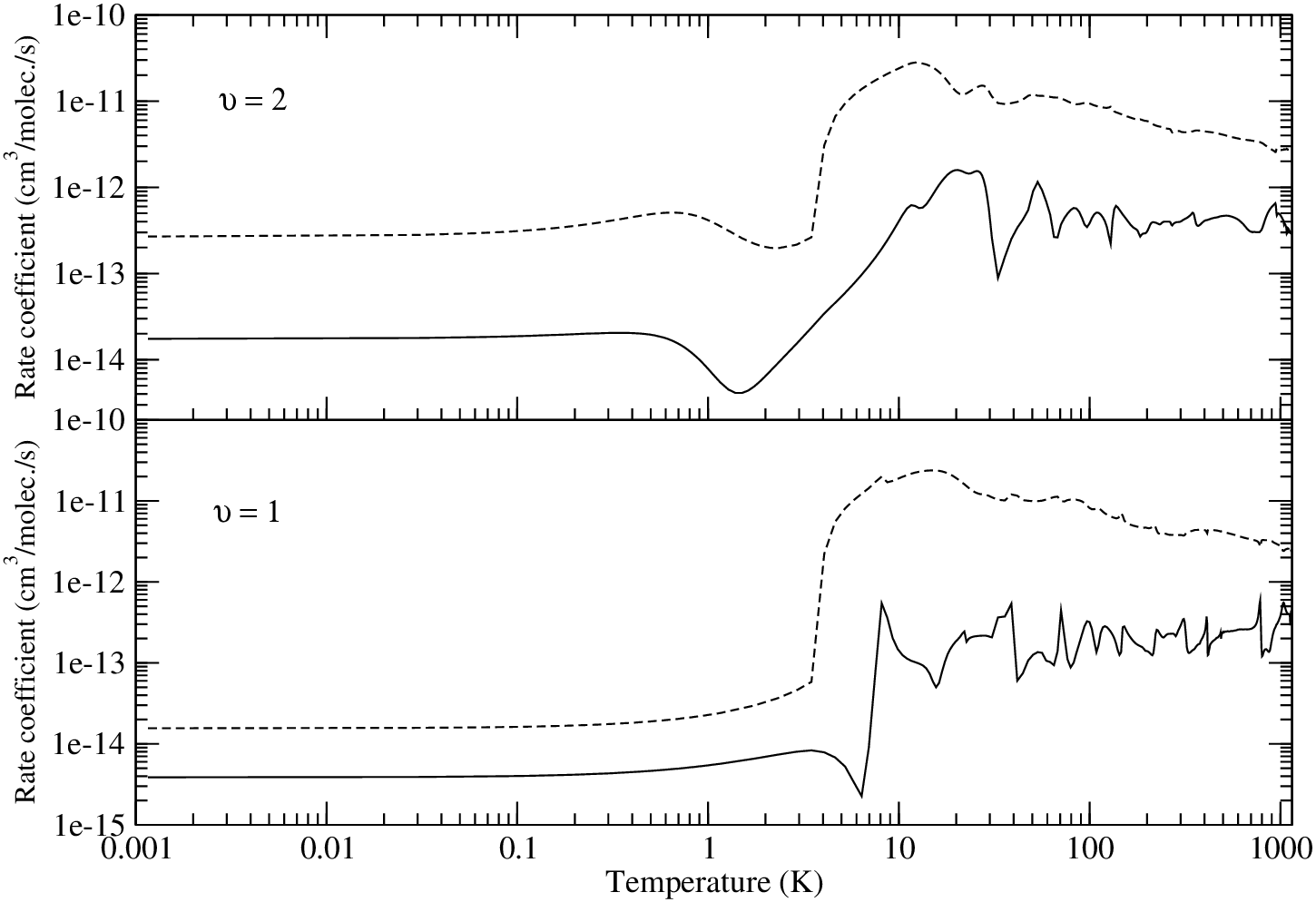}
\end{center} 
\end{figure*}

%=============
\end{document}